\documentclass[%
 reprint,
 amsmath,
 amssymb,
 aps,
 pra,
twocolumn,
]{revtex4-1}

\usepackage{graphicx}
\usepackage{dcolumn}
\usepackage{bm}
\usepackage{gensymb}
\usepackage{physics}
\usepackage{color}
\usepackage{tikz}
\usepackage{soul}
\usepackage{hyperref}

\hypersetup{
    colorlinks=true,       
    linkcolor=blue,          
    citecolor=blue,        
    filecolor=blue,      
    urlcolor=blue           
}

\usepackage{txfonts}

\newcommand{\sr}[1]{\phantom{ }^{#1}\text{Sr}}

\newcommand{\term}[3]{\phantom{ }^{#1}#2_{#3}}

\def\ms{\mbox{ ms}}

\def\us{\mbox{ $\mu$s}}

\begin{document}

\preprint{APS/123-QED}

\title{Enhanced Magnetic Trap Loading for Atomic Strontium}

\author{D.S. Barker}
\email{dbarker2@umd.edu}
\author{B.J. Reschovsky}%
\author{N.C. Pisenti}
\author{G.K. Campbell}
\affiliation{%
 Joint Quantum Institute, University of Maryland and National Institute of Standards and Technology, College Park, MD 20742
}%

\date{\today}

\begin{abstract}
We report on a technique to improve the continuous loading of atomic strontium into a magnetic trap from a Magneto-Optical Trap (MOT). This is achieved by adding a depumping laser tuned to the $\term{3}{P}{1}\rightarrow\term{3}{S}{1}$ (688-nm) transition. The depumping laser increases atom number in the magnetic trap and subsequent cooling stages by up to $65~\%$ for the bosonic isotopes and up to $30~\%$ for the fermionic isotope of strontium. We optimize this trap loading strategy with respect to the 688-nm laser detuning, intensity, and beam size. To understand the results, we develop a one-dimensional rate equation model of the system, which is in good agreement with the data. We discuss the use of other transitions in strontium for accelerated trap loading and the application of the technique to other alkaline-earth-like atoms.


\end{abstract}

\maketitle


\section{\label{intro}Introduction}

Alkaline-earth-like (AE) atoms have received a great deal of recent interest due to the distinctive properties of their level structure~\cite{Sansonetti2010,NIST_ASD}. The largely disconnected singlet and triplet states in these atoms give rise to very narrow optical transitions, which could form the basis for an improved time standard~\cite{Bloom2014,Hinkley2013}. These transitions are also advantageous in a wide variety of other applications. For example, their low photon scattering rates allow for the production of highly-excited Rydberg atoms with reduced decoherence compared to alkali-metals~\cite{DeSalvo2015}. Magnetic field insensitive singlet and triplet levels make AE atoms attractive for precision measurement and quantum sensing applications~\cite{Sorrentino2009,Jamison2014a}. In fermionic isotopes, these states manifest SU($2I+1$) spin symmetry, where $I$ is the nuclear angular momentum, allowing quantum simulation of hamiltonians that are inaccessible with alkali atoms~\cite{Foss-Feig2010, Gorshkov2010, Beverland2014}. All of these applications require or benefit from a combination of large atom number and short experimental cycle times.

\begin{figure}
\includegraphics[width=\linewidth]{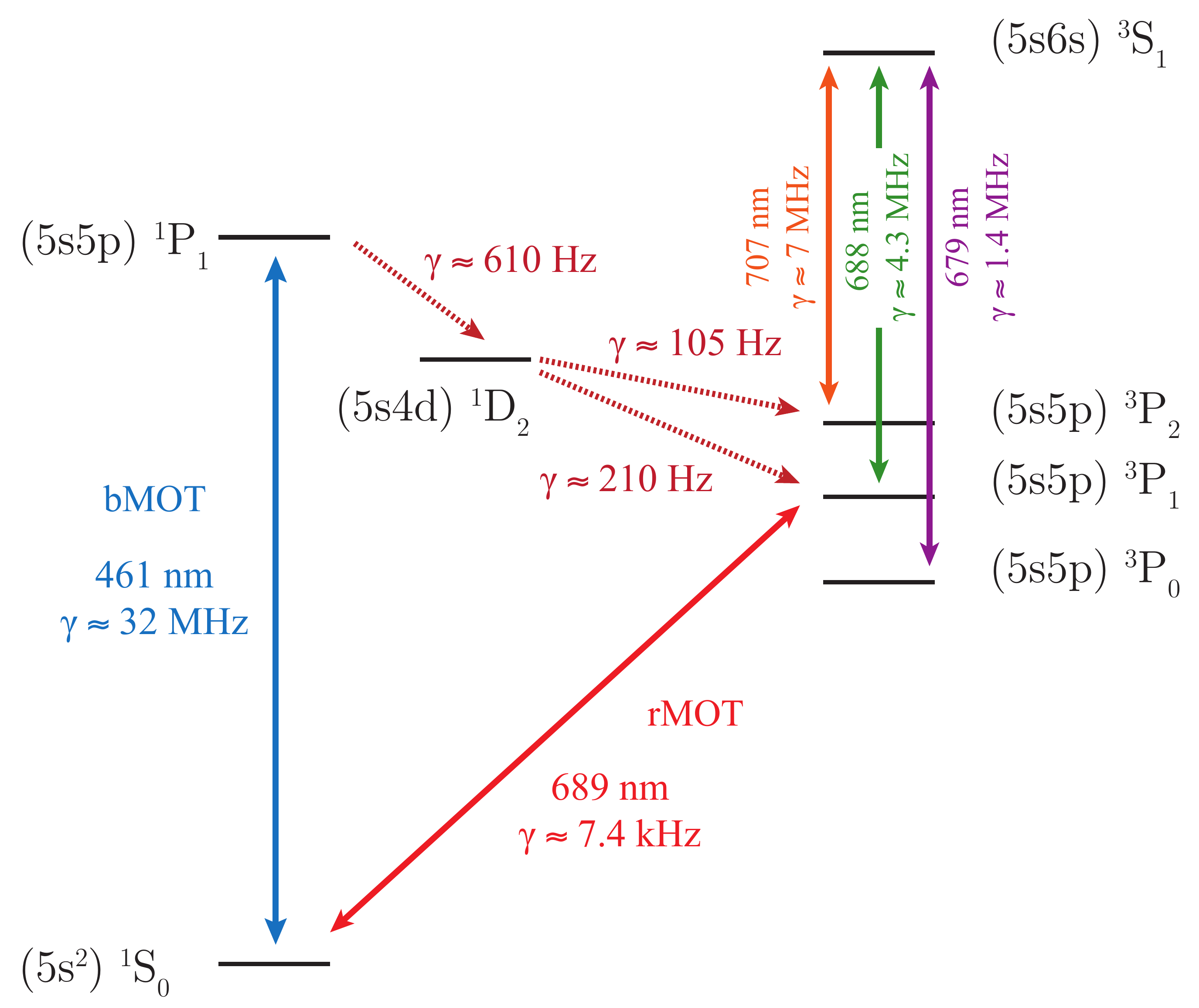}
\caption{(Color online) The low-lying energy levels of bosonic strontium with linewidths taken from~\cite{Bauschlicher1985,Hunter1986,Sansonetti2010}. A Magneto-Optical Trap operating on the blue, 461-nm transition (bMOT) captures atoms from a Zeeman-slowed beam. Atoms in the bMOT continuously leak into the long-lived $^{3}P_{2}$ state, which is magnetically trapped by the bMOT quadrupole field. Two lasers at 688~nm and 679~nm increase the magnetic trap loading rate by pumping atoms that populate $^{3}P_{1}$ into $^{3}P_{2}$. The 679-nm laser and a 707-nm laser return atoms to the ground state via the $^{3}P_{1}$ state once magnetic trap loading is complete. A Magneto-Optical Trap operating on the red, 689-nm transition (rMOT) then cools the sample to ${\approx}1$~$\mu$K.}
\label{levels}
\end{figure}

Recent advances in cooling and trapping techniques enabled production of the first strontium degenerate gases~\cite{DeEscobar2009,Stellmer2009a,DeSalvo2010,Mickelson2010,Stellmer2010}. The small negative s-wave scattering length of the most abundant isotope, $\sr{88}$, hampered initial efforts to create Bose-Einstein condensates~\cite{Ido2000,Ferrari2006,Traverso2008,MartinezdeEscobar2008,Mickelson2009a}. While the other stable isotopes ($\sr{87}$, $\sr{86}$, and $\sr{84}$) possess favorable scattering lengths, their low natural abundance initially prevented Magneto-Optical Traps (MOT) from collecting enough atoms to reach degeneracy. Fortuitously, laser cooling of strontium on the 461-nm line  populates a magnetically-confined, metastable reservoir of atoms in the $\term{3}{P}{2}$ state (see Figure~\ref{levels})~\cite{Nagel2003}. The long lifetime of this reservoir (typically ${\gtrsim}10$~s) compared to the MOT allows for the accumulation of sufficient populations of $\sr{87}$, $\sr{86}$, or $\sr{84}$ for forced evaporation or sympathetic cooling of $^{88}\text{Sr}$~\cite{DeEscobar2009,Stellmer2009a,DeSalvo2010,Mickelson2010,Stellmer2010}. The ${\approx}1$~$\mu$K temperatures attainable with laser cooling on the 689-nm, intercombination transition (Figure~\ref{levels}) lead to short evaporation times to reach degeneracy. Given the low abundance of the interacting isotopes, this means that the reservoir loading time usually dominates the experimental cycle~\cite{DeEscobar2009, Stellmer2013a, Mickelson2010, DeSalvo2010}. 

\begin{figure*}[t]
\includegraphics[width=\linewidth]{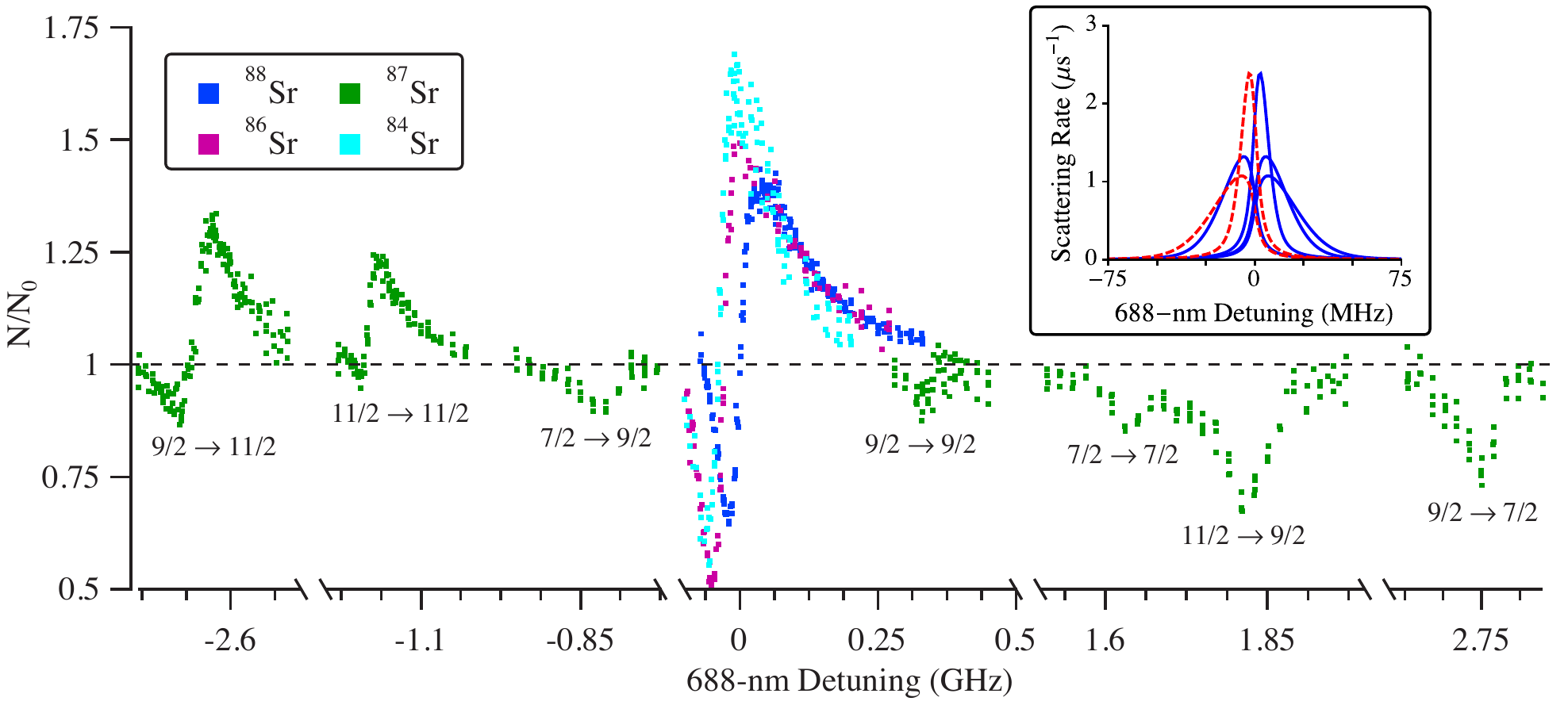}
\caption{(Color online) Atom number enhancement as a function of 688-nm laser detuning, $\Delta_{688}$, for all strontium isotopes. The detuning zero is referenced to $^{88}\text{Sr}$. For the data shown, the rMOT recaptures the less abundant isotopes and the bMOT recaptures $^{88}\text{Sr}$. The depumper saturation parameter was $s_{688}\approx35$ ($\approx50$) for $^{88}\text{Sr}$ ($^{87}\text{Sr}, ^{86}\text{Sr}, ^{84}\text{Sr}$) and the 679-nm laser detuning, $\Delta_{679}$, was set to maximize bMOT fluorescence. We label the fermionic hyperfine transitions with $F\,{\rightarrow}\,F'$, where $F$ is the total angular momentum quantum number for $^{3}P_{1}$ and $F'$ the corresponding quantum number for $^{3}S_{1}$. Inset: The detuning-dependent scattering rate for each transition between $^{3}P_{1}$ and $^{3}S_{1}$ Zeeman levels, averaged over the volume of a one-dimensional bMOT with $s_{688}=1$ (see Section \ref{sim}). Solid blue curves pump to a $^{3}S_{1}$ Zeeman level that can decay to a magnetically trappable $^{3}P_{2}$ Zeeman state, but dashed red curves do not. Asymmetric lineshapes arise in the atom number enhancement because the dashed red scattering rate curves dominate at negative detuning.}
\label{detscan}
\end{figure*}

Typical Sr degenerate gas experiments first use a MOT operating on the $\term{1}{S}{0}\rightarrow\term{1}{P}{1}$, 461-nm transition (bMOT) to capture atoms from a Zeeman-slowed atomic beam and cool them to ${\approx}1$~mK. Atoms slowly leak out of the bMOT cooling cycle (1:50,000 branching ratio) and into the metastable $^{3}P$ manifold, where they populate the $\term{3}{P}{2}$ and $\term{3}{P}{1}$ states in a 1:2 ratio~\cite{Bauschlicher1985,Hunter1986}. The bMOT quadrupole field can magnetically trap atoms in the $\term{3}{P}{2}$ state (the Land\'{e} $g$-factor $g_{J}=3/2$ for bosonic isotopes, where $J$ is the electronic angular momentum). Repumping lasers return $\term{3}{P}{2}$ atoms to the ground state once magnetic trap loading is complete, which, depending on the isotope, can take 30~s or more~\cite{DeEscobar2009,DeSalvo2010,Stellmer2013a}. Loading times are also long for experiments with isotopic mixtures, since the isotope shifts of the 461-nm transition are on the same order of magnitude as the linewidth~\cite{Mickelson2010,Stellmer2013a}. This prohibits efficient simultaneous loading of the magnetic trap. A second stage Magneto-Optical Trap using the 689-nm, intercombination line (rMOT) cools these atoms to ${\approx}1$~$\mu$K and facilitates loading into an optical dipole trap. Evaporation proceeds quickly due to the low initial temperature and degeneracy can be reached in ${\approx}1$~s for most isotopes~\cite{Stellmer2013a}.

Here we present a technique to reduce the reservoir loading time or, equivalently, increase the atom number for experiments with strontium as first suggested in~\cite{Stellmer2013}. The method relies on continuous optical pumping of atoms from the short-lived $\term{3}{P}{1}$ state into the magnetically trapped $\term{3}{P}{2}$ reservoir using the $\term{3}{P}{1}\rightarrow\term{3}{S}{1}$, 688-nm transition. This greatly reduces the steady-state atom number in the bMOT, but increases the flux of low-field seeking atoms into the metastable reservoir. Although the $\term{3}{P}{2}{:}\term{3}{P}{1}$ branching ratio from the $\term{1}{D}{2}$ state suggests that atom number should be enhanced by a factor of three (see Figure~\ref{levels}), we show that this estimate is incorrect since it does not consider the reduction in bMOT atom number caused by the 688-nm laser.

We describe our experimental apparatus in Section \ref{exp} with an emphasis on the details relevant for the accelerated loading scheme. Section \ref{res} explains the measurement procedure and results. In Section \ref{sim}, we develop a rate equation model and demonstrate that our data is in agreement with expectations. We also simulate the trap loading enhancement for several other transitions in strontium and two in calcium. Section \ref{con} is a summary of our results and outlook.

\section{\label{exp}Apparatus}

Our experimental setup is similar to other strontium apparatuses designed for optical clock and degenerate gas experiments~\cite{Stellmer2013,Mickelson2010a,Boyd2007,Ludlow2008}. An oven with a microtubule array nozzle, heated to 600~$\degree$C, creates an atomic strontium beam. Two stages of differential pumping prevent the pressure in the experiment chamber ($6\times10^{-11}$~Torr) from rising while the oven is in operation. The atomic beam passes through a transverse cooling stage, which consists of two orthogonal, retroreflected 461-nm laser beams. Each beam has ${\approx}10$~mW of power, a 1:3 aspect ratio ($1/e^{2}$ radius of 9~mm along the atomic beam axis), and $-10$~MHz detuning from the $\term{1}{S}{0}\rightarrow\term{1}{P}{1}$ transition. The Zeeman slower is a 35-cm long, multilayer, variable pitch coil located immediately after the transverse cooling stage. It is pumped with ${\approx}48$~mW of $-600$~MHz detuned 461-nm light, which is focused onto the oven nozzle with an initial $1/e^{2}$ radius of 5~mm.

The bMOT has a standard retroreflected, six beam configuration. Each beam has a $1/e^{2}$ radius of 8~mm, a detuning $\Delta_{461}= -45$~MHz, and contains either ${\approx}7$~mW (for bosonic isotope data) or ${\approx}9$~mW (for $\sr{87}$ data) of power. These parameters give $s_{461}=I/I_{sat}\approx0.16$ per beam for the bosons and $s_{461}\approx0.21$ per beam for the fermion. The quadrupole coil has a vertical axis of symmetry and produces a magnetic field gradient of 6~mT/cm along that axis during bMOT operation. The bMOT field gradient is sufficient to magnetically trap $\term{3}{P}{2}$ atoms in the low-field seeking $\ket{m_{J}=1}$ and $\ket{m_{J}=2}$ Zeeman sublevels. In our vacuum chamber, the position of two recessed viewports along the symmetry axis of the coils limits the trap depth for the $\ket{m_{J}=1}$ state to ${\approx}5$~mK. This limitation is unimportant for us since our bMOT loads ${\approx}1$~mK atoms into the magnetic trap, but it suggests that experiments with larger vacuum chambers may find that a higher temperature bMOT optimizes magnetic trap loading~\cite{Stellmer2013}.

\begin{figure}[t]
\includegraphics[width=\linewidth]{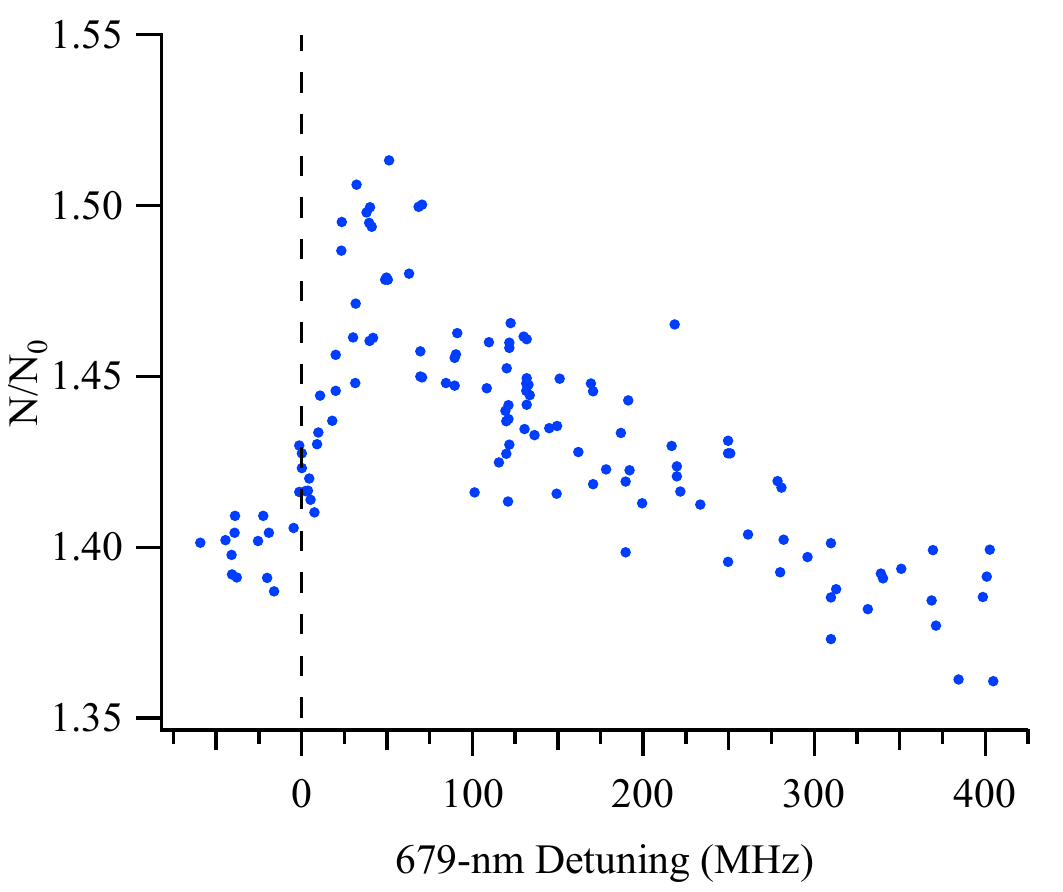}
\caption{(Color online) $N/N_{0}$ for $^{88}\text{Sr}$ recaptured in the bMOT as a function of $\Delta_{679}$. The 688-nm laser has $s_{688}\approx35$ and $\Delta_{688}\approx30$~MHz. We reference the 679-nm detuning to the bMOT fluorescence maximum, indicated by the dashed vertical line. As shown in the inset to Figure~\ref{detscan}, the asymmetry arises from detuning-dependent scattering rates. Transitions to $^{3}S_{1}$ Zeeman levels that can decay into the magnetic trap are predominately blue detuned, whereas red detuned transitions populate levels that decay to high-field seeking $^{3}P_{2}$ Zeeman states.}
\label{det679}
\end{figure}

Two repumping lasers addressing the 679-nm, $\term{3}{P}{0}\rightarrow\term{3}{S}{1}$ and the 707-nm, $\term{3}{P}{2}\rightarrow\term{3}{S}{1}$ transitions are used to return $\term{3}{P}{2}$ atoms in the magnetic trap to the ground state. The two beams co-propagate with the Zeeman slower beam, share a $1/e^{2}$ radius of ${\approx}1$~cm, and contain ${\approx}2.5$~mW (679~nm) and ${\approx}4.5$~mW (707~nm) of power. For experiments with the bosonic isotopes, we lock the repump laser frequencies using slow feedback from a HighFinesse WS7 wavemeter~\footnote{The identification of commercial products is for information only and does not imply recommendation or endorsement by the National Institute of Standards and Technology.}. The locking stability is $\pm5$~MHz, which is much narrower than the observed bosonic repumping linewidth~\footnote{The WS7 wavemeter resolution is specified to be 10~MHz, but we find that it can reliably detect 1~MHz frequency offsets.}. The presence of hyperfine structure in the fermion complicates repumping on the $\term{3}{P}{2}\rightarrow\term{3}{S}{1}$ transition. In order to cover as much of the ${\approx}5.5$~GHz hyperfine spectrum of the transition as possible, we modulate the 707-nm laser frequency at ${\approx}700$~Hz. To further increase coverage of the hyperfine spectrum, we use a second 707-nm laser that we modulate at ${\approx}600$~Hz. When optimized, application of the second laser to the experiment increases the $\sr{87}$ atom number by about $10~\%$. For the fermionic data, the 679-nm laser is locked to the $|\term{3}{P}{0},F=9/2\rangle\rightarrow|\term{3}{S}{1},F=11/2\rangle$ transition (where $F=I+J$) using the wavemeter.

The linewidth of our 689-nm master oscillator is stabilized below the natural linewidth of the $\term{1}{S}{0}\rightarrow\term{3}{P}{1}$ resonance using a Pound-Drever-Hall lock to an optical cavity (finesse ${\approx}240,000$)~\cite{Drever1983}. We injection lock a slave laser diode to the master to obtain sufficient power for trapping (for the fermion, this laser pumps the $|\term{1}{S}{0},F=9/2\rangle\rightarrow|\term{3}{P}{1},F=11/2\rangle$ transition). Dichroic beamsplitters overlap the 689-nm light for the rMOT with the bMOT beams. In each rMOT arm, the power is ${\approx}3.5$~mW and the $1/e^{2}$ radius is 2.5~mm. The quadrupole field gradient switches to 0.16~mT/cm for rMOT operation. For the first 100~ms of rMOT operation, we frequency modulate the trapping laser at 30~kHz with a modulation depth of 1~MHz to increase the capture velocity of the rMOT. Over the next 400~ms, we linearly reduce the modulation depth to 100~kHz while simultaneously ramping the optical power to 100~$\mu$W with a half-gaussian temporal profile. In this work, we terminate rMOT operation at this stage ($T\approx2$~$\mu$K), but we can cool further by turning off the frequency modulation and reducing the intensity. The hyperfine structure of the fermion means that a second slave laser is required to make a stable rMOT~\cite{Mukaiyama2003}. A beatnote lock to the master stabilizes the second laser to the $|\term{1}{S}{0},F=9/2\rangle\rightarrow|\term{3}{P}{1},F=9/2\rangle$ line~\cite{Appel2008}. The second slave laser provides $\approx800$~$\mu$W of light per rMOT beam to the experiment. Aside from the reduction in initial power, the intensity and modulation ramps are identical to those for the trapping laser. We also linearly increase the rMOT magnetic field gradient to 0.24~mT/cm during the final 400~ms of the rMOT when trapping $\sr{87}$ to increase the atomic density.

To enhance magnetic trap loading, a 688-nm laser resonant with the $\term{3}{P}{1}\rightarrow\term{3}{S}{1}$ transition pumps atoms that decay to $\term{3}{P}{1}$ out of the bMOT cycle and into $\term{3}{P}{2}$. We call this laser the depumper because it makes the bMOT transition less closed. The depumper is a Littman-Metcalf configuration laser which we built using a laser diode (model HL6738MG~\footnotemark[1]) that was AR-coated in-house~\footnote{The anti-reflection coating is comprised of two layers, as suggested in \cite{Fox1997}:  one layer of $\text{Al}_{2}\text{O}_{3}$ to bring the facet coating to $\lambda/2$, and a final $\lambda/4$ layer of $\text{HfO}_{2}$. The deposition was done via electron beam evaporation, and monitored in-situ by scanning the diode current across the lasing threshold.}. The laser provides up to 2.8~mW of light to the experiment, which corresponds to $s_{688}=I/I_{sat}\approx50$. The 688-nm beam enters the chamber horizontally and perpendicular to the Zeeman slower axis. It has a $1/e^{2}$ radius $w_{688}=1.35$~mm except where otherwise noted. We stabilize the 688-nm laser detuning, $\Delta_{688}$, to within $\pm3$~MHz by locking to the wavemeter.

We measure the magnetic trap loading enhancement by interleaving shots with the 688-nm laser on and off. The \mbox{679-nm} repumping laser closes the $\term{3}{S}{1}\rightarrow\term{3}{P}{0}$ leak to increase the depumper's effect. It remains on during both shots of a depumper on/off pair of experimental runs, but does not affect trap loading when the 688-nm laser is off since the bMOT does not populate $\term{3}{P}{0}$. After $0.5$~s to $30$~s of reservoir loading, an acousto-optic modulator extinguishes the 688-nm beam and optical shutters open to allow 707-nm light to reach the experiment. Either the bMOT or the rMOT can recapture atoms from the magnetic trap for detection and imaging. However, rMOT recapture greatly improves signal-to-noise for $\sr{87}$, $\sr{86}$, and $\sr{84}$, so we use it exclusively for these isotopes. The bMOT recapture stage lasts 100~ms and rMOT recapture consists of the full rMOT cycle described above. We take an absorption image after a 1~ms (25~ms) ballistic expansion for bMOT (rMOT) recapture using a resonant, 10$\us$ pulse of 461-nm light with $I/I_{sat}\approx0.04$. Numerical integration of the image yields the atom number for each shot.

\section{\label{res}Results}

\begin{figure}[t]
\includegraphics[width=\linewidth]{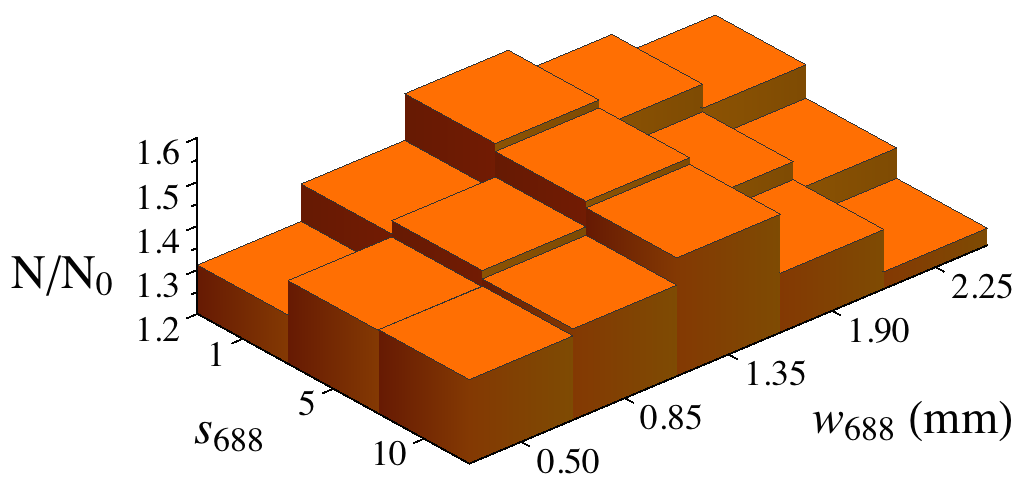}
\caption{(Color online) $N/N_{0}$ for $^{88}\text{Sr}$ recaptured in the bMOT versus $s_{688}$ and $w_{688}$, where $w_{688}$ is the $1/e^{2}$ radius of the depump beam. The standard errors are omitted for clarity, but are $\lesssim 0.03$. The optimal trap enhancement occurs when $w_{688}$ roughly matches the $1/e$ radius of the bMOT and $s_{688}\approx1$. $\Delta_{688}$ and $\Delta_{679}$ are both ${\approx}30$~MHz (corresponding to the maxima in Figures~\ref{detscan} and~\ref{det679}).}
\label{sizesat}
\end{figure}

We study the magnetic trap loading enhancement as a function of isotope, power, detuning, and beam size. The enhancement is measured by comparing the atom number recaptured in the rMOT or bMOT with and without the 688-nm laser. We find that the depumper's effect is independent of which MOT we use for atom recapture. The magnetic trap loading enhancement is given by the normalized atom number, $N/N_{0}$, where $N$ is the atom number with the depumper on and $N_{0}$ the number with it off. 

We investigate the loading enhancement as we scan the depumper across the 688-nm transition. For this data set, we set $s_{688}\approx35$ for $\sr{88}$ and $s_{688}\approx50$ for all other isotopes. The repump laser frequencies are locked to maximize bMOT fluorescence. The magnetic trap loading time, $t_{\text{load}}$, for \{$\sr{88}$, $\sr{87}$, $\sr{86}$, $\sr{84}$\} is \{1.5~s, 10~s, 6~s, 7.5~s\} resulting in typical $N_{0}$ of $\{2\times10^{7},5\times10^{6},1\times10^{7},8\times10^{5}\}$ in the rMOT. Adjustment of the wavemeter lockpoint allows us to scan $\Delta_{688}$. Figure~\ref{detscan} shows the depumping spectrum for all isotopes and hyperfine transitions, the locations of which are in good agreement with~\cite{Courtillot2005}. On average, we observe trap loading improvements of ${\approx}50~\%$ for bosonic isotopes and ${\approx}25~\%$ for the fermionic isotope even without detailed optimization of the depumping parameters. The peak enhancement for each isotope occurs when $\Delta_{688}\approx30$~MHz from resonance. In 
Figure~\ref{detscan} we also see that the choice of hyperfine transition is crucial for atom number gains in $\sr{87}$. Pumping to $|\term{3}{S}{1},F\,{=}\,7/2\rangle$ and $|\term{3}{S}{1},F\,{=}\,9/2\rangle$ is always detrimental because these manifolds decay with ${\geq}60~\%$ probability to $|\term{3}{P}{2},F\,{=}\,7/2\rangle$ and $|\term{3}{P}{2},F\,{=}\,9/2\rangle$, which have Land\'{e} $g$-factors too small for magnetic trapping at the bMOT field gradient. Pumping to $|\term{3}{S}{1},F\,{=}\,11/2\rangle$ yields a lineshape similar to that of bosonic isotopes, but with reduced amplitude.

\begin{figure}[b]
\includegraphics[width=\linewidth]{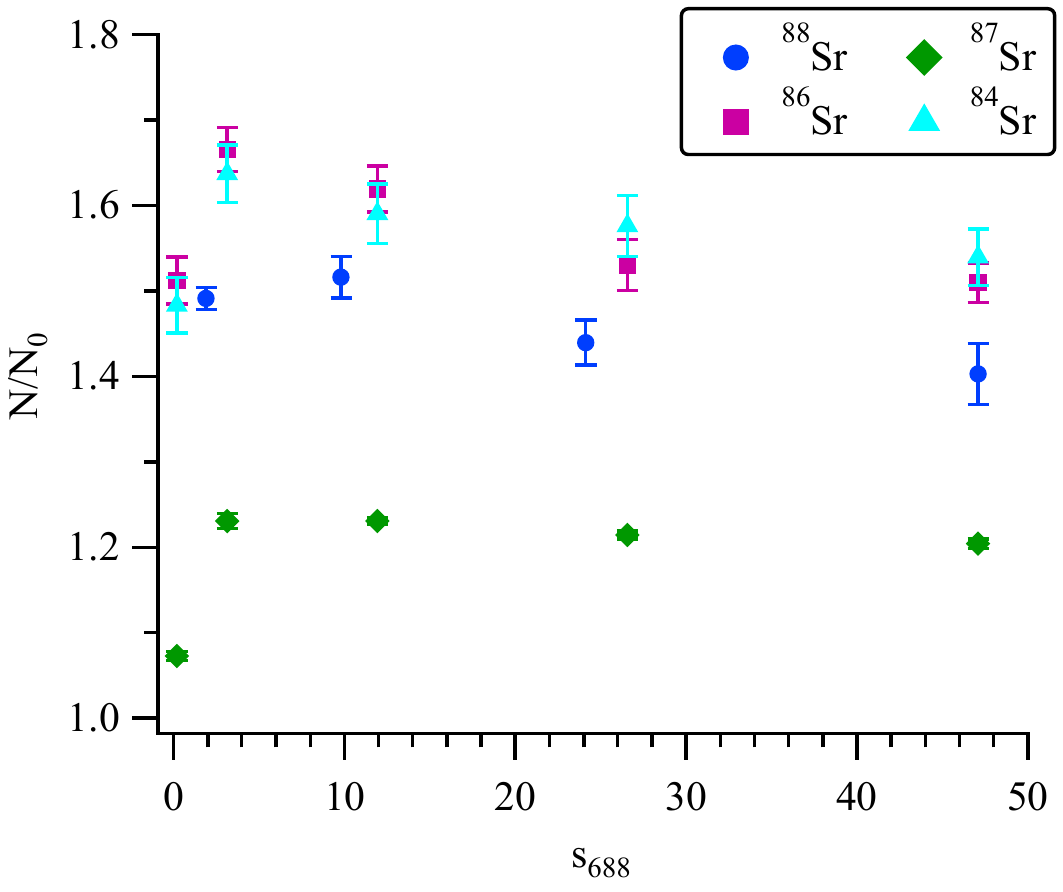}
\caption{(Color online) Atom number enhancement versus $s_{688}$ with $w_{688}=1.35$~mm. $^{87}\text{Sr}$ data are for the $F\,{=}\,11/2 \rightarrow F'\,{=}\,11/2$ transition and all isotopes are recaptured in the rMOT. Error bars represent the standard error in the mean for ${\geq}\,10$ measurements. The 688-nm and 679-nm laser detunings are set to $\approx30$~MHz.}
\label{sat688}
\end{figure}

The asymmetric lineshapes observed in Figure~\ref{detscan} are due to the non-uniform magnetic fields in the bMOT. Quadrupole fields shift low-field seeking states to higher energy and high-field seeking states to lower energy. Because the Land\'{e} $g$-factor for $\term{3}{S}{1}$ is larger than for $\term{3}{P}{1}$, this effect causes a blueshift for most transitions to $\term{3}{S}{1}$ states that can decay to $|\term{3}{P}{2}, m_{J}\,{=}\,1\rangle$ or $|\term{3}{P}{2}, m_{J}\,{=}\,2\rangle$ (see inset to Figure~\ref{detscan}). For the same reason, transitions to $\term{3}{S}{1}$ states that can only populate untrapped $\term{3}{P}{2}$ Zeeman states are redshifted. As a result, trap loading is enhanced to the blue of resonance and reduced to the red of resonance.

For this depumping scheme, application of the 679-nm laser during the bMOT is crucial; removing the 679-nm laser results in ${\approx}50~\%$ reduction of the enhancement. The effect of the 679-nm laser was studied by varying its detuning, $\Delta_{679}$. As shown in Figure~\ref{det679}, setting $\Delta_{679}\approx30$~MHz (relative to the detuning that maximizes the bMOT fluorescence) adds an additional ${\approx}10~\%$ to the enhancement. The asymmetric lineshape is caused by the same mechanism discussed above for the 688-nm transition.

With $\Delta_{688}$ and $\Delta_{679}$ stabilized at their optimized values, we study the trap loading enhancement for $\sr{88}$ as a function of $s_{688}$ and $w_{688}$. Slight focusing/defocusing of the 688-nm beam changes the waist at the location of the bMOT, but the Rayleigh range is always larger than the bMOT $1/e$ radius, $r_{\text{bMOT}}$, for the parameter range we study. Figure~\ref{sizesat} shows that trap loading enhancement increases with $s_{688}$ provided $w_{688}\lesssim r_{\text{bMOT}}$, with the optimal enhancement occurring when $w_{688}\simeq r_{\text{bMOT}}$. High $s_{688}$ increasingly reduces $N/N_{0}$ for larger beam waists. The data suggest that, for our bMOT parameters, a substantial number of atoms populate the $^{3}P$ manifold before being fully captured by the bMOT. These atoms exist outside the bMOT radius and are too hot for magnetic confinement, but they are cold enough that they do not leave the bMOT capture volume during the ${\approx}1$~ms decay time for the $\term{1}{P}{1}\,{\rightarrow}\,\term{1}{D}{2}\,{\rightarrow}\,\term{3}{P}{1}\,{\rightarrow}\,\term{1}{S}{0}$ path. The effect of varying $s_{688}$ and $w_{688}$ in the other isotopes was similar to the $\sr{88}$ results. In Figure~\ref{sat688}, we plot $N/N_{0}$ for a wider range of the saturation parameter at the optimum $w_{688}$. All isotopes exhibit a steep rise in trap loading enhancement for $s_{688}\lesssim1$, followed by a shallow rolloff for $s_{688}>1$. We find that the enhancement is very sensitive to the 688-nm beam alignment and that the peak at $s_{688}\approx1$ is present only when the beam traverses the center of the bMOT.

Before recapturing atoms from the magnetic trap, we do not first discard ground state atoms remaining in the bMOT. This increases both $N$ and $N_{0}$, but decreases their ratio. This choice biases our results toward lower enhancement values, particularly for short load times and for $\sr{88}$. However, the reduced $N/N_{0}$ is the appropriate metric for evaluating the loading enhancement in most experiments, since the cycle time is typically limited by $N$. In situations where bMOT atoms are lost before recapturing from the magnetic trap, which arise in experiments with isotopic mixtures, the depumping technique is even more useful. If we remove the bMOT atoms before imaging, $N/N_{0}$ increases by up to $15~\%$.

\begin{figure}[h]
\includegraphics[width=\linewidth]{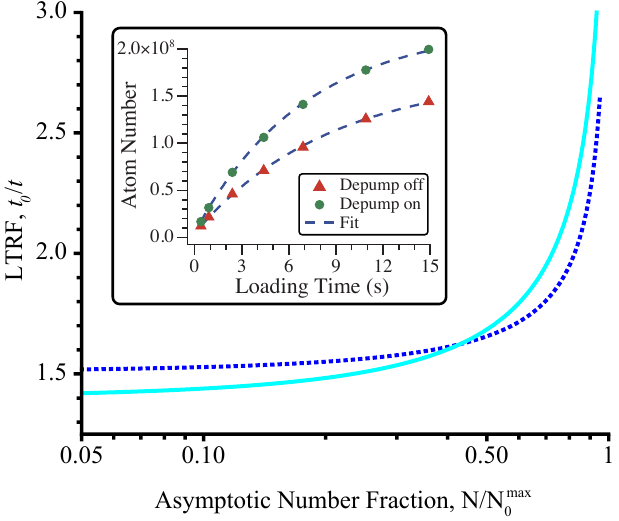}
\caption{(Color online) The Loading Time Reduction Factor (LTRF) for $^{88}\text{Sr}$ (dotted blue line) and $^{84}\text{Sr}$ (solid cyan line). The saturated atom number, $N_{0}^{\text{max}}$, is the asymptote of the trap loading curve with the 688-nm laser off (for $^{88}\text{Sr}$, red triangles, inset). The loading time necessary to transfer $N$ atoms into the rMOT with the depump laser on (off) is given by $t$ ($t_{0}$). Inset: The raw data and fits associated with the $^{88}\text{Sr}$ (dotted blue) LTRF curve. The standard error is smaller than the data points. We find $t$ and $t_{0}$ by inverting the appropriate fit function (see text).}
\label{LTR}
\end{figure}

Our technique reduces the trap loading time necessary to achieve a given atom number. For short trap loading times, $N/N_{0}$ is a measure of the increased loading rate achieved with the depumping laser. This is the regime shown in Figure~\ref{detscan}. For longer trap loading times, the atom number will saturate. Experiments requiring very large atom numbers close to the saturation limit can expect even greater reductions in loading time than suggested by the initial loading rate. We demonstrate this by fitting $N(t)$ and $N_{0}(t)$ with $N(t)\,{=}\,N^{\text{max}}(1-e^{-\alpha t})$, where $\alpha$ is the loading time constant and $N^{\text{max}}$ is the saturated atom number (see inset to Figure~\ref{LTR}). Inverting the fitted function yields the loading time necessary to reach a given atom number with the depumper on, $t(N)$, or with the depumper off, $t_{0}(N_{0})$. We plot the loading time reduction factor, $\text{LTRF}(N/N_{0}^{\text{max}})=t_{0}(N_{0}\text{=}N)/t(N)$, for $\sr{88}$ and $\sr{84}$ in Figure~\ref{LTR}. The loading time reduction diverges as $N\rightarrow N_{0}^{\text{max}}$ since $N^{\text{max}} > N_{0}^{\text{max}}$. For example, to reach an atom number of ${\approx} N_{0}^{\text{max}}$, the depumping technique can reduce the loading time by a factor of ${\approx}3$.

\section{\label{sim}Simulation}

To better understand the enhancement, we develop a one-dimensional rate equation model to simulate the bMOT depumping process. This was motivated by two features of our data: the asymmetric lineshapes depicted in Figure~\ref{detscan}, and the discrepancy between the observed performance and the $3\times$ initial estimate given by the $\term{3}{P}{2}{:}\term{3}{P}{1}$ branching ratio. A simple calculation, based on analysis of the cascade of Clebsch-Gordan (CG) coefficients connecting $\term{1}{P}{1}$ to $\term{3}{P}{1}$ and $\term{3}{P}{2}$, suggests that $3\times$ trap loading enhancement is unlikely. However, this CG calculation depends sensitively on the relative populations of the $\term{1}{P}{1}$ Zeeman sublevels, which are position dependent, and the steady state atom number in the bMOT. Both of these complications prevent analysis of experimental performance by this method. A full simulation of the optical pumping dynamics resolves both of these issues, allowing direct comparison of data with theory. Straightforward modifications of the rate equation model allow us to compare our technique to alternative depumping transitions.

In the rate equation model, we track the population, $P_{i,m_{i}}$, in each magnetic sublevel of $i\in\{ \term{1}{S}{0}, \term{1}{P}{1}, \ldots, \term{3}{S}{1} \}$, with $m_{i}$ the spin projection along the axis of a one-dimensional bMOT. Each level decays at a rate given by the appropriate linewidth, $\gamma_{ij}$, from Figure~\ref{levels}, 
\begin{equation}
\Gamma^{decay}_{\ket{i,m_{i}}\rightarrow |{j,m_{j};\ m_{\gamma}}\rangle}=\gamma_{ij}\left|\bra{j, m_{j}; 1, m_{\gamma}}\ket{ i, m_{i}}\right|^{2}\,,
\end{equation}
where $\langle i, m_{i}; 1, m_{\gamma}| j, m_{j}\rangle$ is the CG coefficient. In addition to the transitions shown in Figure~\ref{levels}, we also include the $\term{1}{D}{2}\rightarrow\term{1}{S}{0}$ quadrupole decay because its linewidth is non-negligible compared to decay rates into the $^{3}P$ states~\cite{Bauschlicher1985}. Since we are not interested in individual atom trajectories, we average the driven excitation rate for $\ket{i,m_{i}}\rightarrow |{j,m_{j}}\rangle$, $\Gamma^{exc}_{|{i,m_{i};\ m_{\gamma}}\rangle\rightarrow |{j,m_{j}}\rangle}$, over the position and velocity distribution of the MOT,
\begin{equation}
\rho(x,v)=\frac{e^{-mv^{2}/2 k_{b} T}e^{-(x/r_{\text{bMOT}})^{2}}}{\pi r^{2}_{\text{bMOT}}\sqrt{2\pi k_{B} T/m}}\,.
\end{equation}
For the $\term{3}{P}{0}\rightarrow\term{3}{S}{1}$ and $\term{3}{P}{1}\rightarrow\term{3}{S}{1}$ transitions, we arrive at
\begin{equation}
\label{scats}
\begin{split}
& \Gamma^{exc}_{|{i,m_{i};\ m_{\gamma}}\rangle\rightarrow |{j,m_{j}}\rangle} = \\ 
& {\iint\limits_{0,\ -v_{max}}^{x_{max},\ v_{max}}}\rho(x,v)\frac{ s_{ij}\ \gamma_{ij}\ \sigma(m_{\gamma})\left|\bra{i, m_{i}; 1, m_{\gamma}}\ket{ j, m_{j}}\right|^{2}}{1+s_{ij}+4(\Delta_{m_{i}m_{j}}/\gamma_{ij})^{2}}dx\,dv\,,
\end{split}
\end{equation}
where $s_{ij}$ is the saturation parameter, $\sigma(m_{\gamma})$ is the fraction of $s_{ij}$ with polarization $m_{\gamma}\in\{-1,0,1\}$, and the effective detuning between $\ket{i,m_{i}}$ and $|{j,m_{j}}\rangle$, $\Delta_{m_{i}m_{j}}$, includes Doppler and Zeeman shifts. We choose $x_{max}$, $v_{max}$ to be much larger than the characteristic scale of $\rho(x,v)$. The symmetry of a one-dimensional MOT permits us to model only the $x>0$ region with all scattering rates then multiplied by two. This choice simplifies the tracking of magnetically trapped atoms because the magnetic field does not change sign in the simulation volume. Taking the transformation $s_{ij}\rightarrow2 s_{ij}$ in Equation (\ref{scats}) while maintaining $\Sigma_{m_{\gamma}} \sigma(m_{\gamma})=1$ gives the correct scattering rate for the two bMOT beams. We assume a pure circular polarization for both bMOT beams and a random polarization for the repumper and depumper.

\begin{figure}[t]
\includegraphics[width=\linewidth]{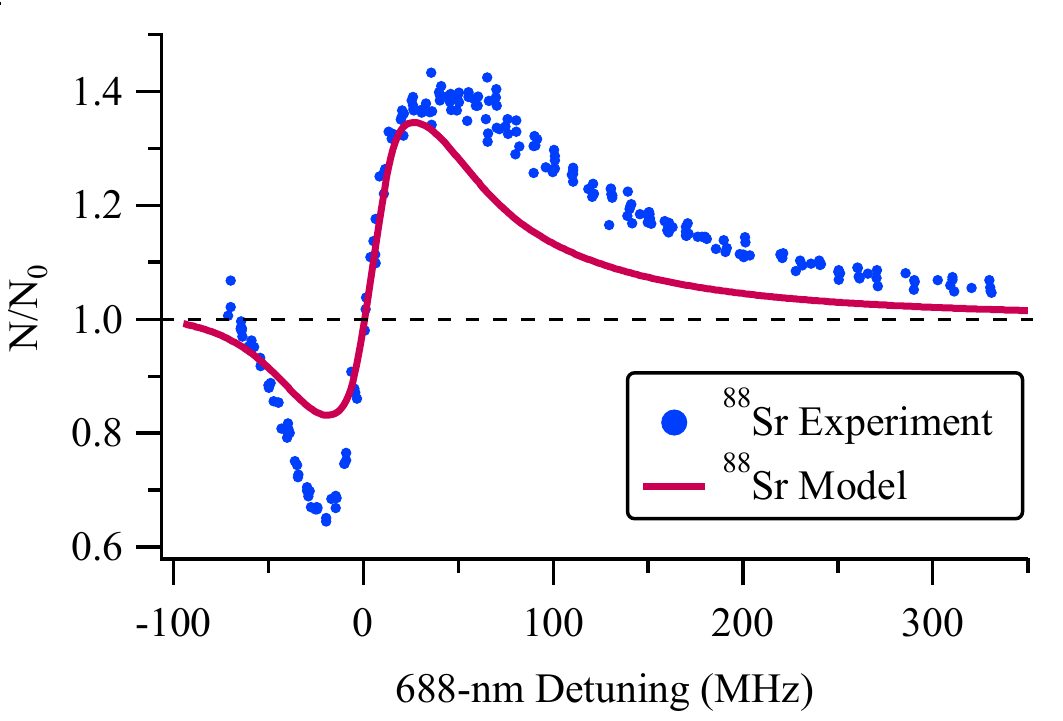}
\caption{(Color online) The $^{88}\text{Sr}$ curve from Figure~\ref{detscan} (blue circles) plotted with simulation results (magenta line) from the one-dimensional rate equation model described in Section \ref{sim}. Parameters for the simulation are identical to experimental conditions (see Section \ref{exp}). The model reproduces the qualitative lineshape and the peak enhancement of the data. The difference in the amplitude and width between the curves implies that our one-dimensional model does not accurately capture the full three-dimensional nature of our experiment.}
\label{SimvExp}
\end{figure}

We describe the evolution of the populations, $\{P_{i,m_{i}}\}$, with a system of coupled differential equations,
\begin{equation}
\begin{split}
&\dot P_{i,m_{i}} = \mathcal{F}_{\term{1}{S}{0}}\delta_{i,\term{1}{S}{0}} - \beta (\delta_{i,\term{1}{S}{0}}+\delta_{i,\term{3}{P}{2}}) P_{i,m_{i}} \\ 
&+ {\sum_{j,m_{j},m_{\gamma}}}\Big( \Gamma^{exc}_{|{j,m_{j};\ m_{\gamma}}\rangle\rightarrow |{i,m_{i}}\rangle}P_{j,m_{j}} - \Gamma^{exc}_{|{i,m_{i};\ m_{\gamma}}\rangle\rightarrow |{j,m_{j}}\rangle}P_{i,m_{i}} \\
&+ \Gamma^{decay}_{|{j,m_{j}}\rangle\rightarrow |{i,m_{i};\ m_{\gamma}}\rangle}P_{j,m_{j}} - \Gamma^{decay}_{\ket{i,m_{i}}\rightarrow |{j,m_{j};\ m_{\gamma}}\rangle}P_{i,m_{i}}\Big)\,,
\end{split}
\end{equation}
where $\delta_{i,j}$ is the Kronecker delta, $\mathcal{F}_{\term{1}{S}{0}}$ is the atomic flux from the Zeeman slower, and $\beta\,{\approx}\,0.1~\text{s}^{-1}$ is the experimentally measured 1-body loss rate (the effect of which is negligible for states with short lifetimes). Without repumping, the bMOT loading time (${\lesssim}\,100\ms$) is short compared to the magnetic trap loading time, so we take $\dot P_{i,m_{i}}=0$ for all $i \ne \term{3}{P}{2}$. We solve algebraically for $\{\dot P_{\term{3}{P}{2},-2},\ldots,\dot P_{\term{3}{P}{2},2}\}$ and numerically integrate the resulting first-order equations from $t=0$ to $t=t_{\text{load}}$. The sum $(P_{\term{3}{P}{2},2}+P_{\term{3}{P}{2},1}+P_{\term{1}{S}{0},0})$ gives the total atom number at $t=t_{\text{load}}$ (the population of other states is negligible), which we equate with $N$ or $N_{0}$ depending on whether the 688-nm laser is on or off. A fit of the model to the $\sr{88}$, $N_{0}$ versus $t_{\text{load}}$ data, with $s_{\term{3}{P}{1}\term{3}{S}{1}}\equiv s_{688} = 0$ and $\mathcal{F}_{\term{1}{S}{0}}$ as the only free parameter, matches the experiment to better than $4~\%$ for all reservoir loading times (all other parameters are taken from Section \ref{exp}). We use the extracted value of $\mathcal{F}_{\term{1}{S}{0}}$ for all subsequent simulations, but we find that the results are independent of $\mathcal{F}_{\term{1}{S}{0}}$ and $\beta$.

We plot the simulated and measured $N/N_{0}$ for $\sr{88}$ in Figure~\ref{SimvExp}. Parameters for the simulation are taken from Section \ref{exp} except for $s_{688}$ and $\Delta_{679}$, which are the same as given in Figure~\ref{detscan}. The simulation agrees reasonably well with experiment given the simplicity of the model and the absence of free parameters. It approximately reproduces the asymmetric lineshape and the magnitude of the peak trap loading enhancement. The one-dimensional model also qualitatively replicates the behavior of $N/N_{0}$ as a function of $s_{688}$ and $\Delta_{679}$. The difference in dimensionality between the 1D simulation and 3D experiment causes the mismatch in both the width and amplitude of the lineshapes in Figure~\ref{SimvExp}. The three-dimensional MOT beam configuration and magnetic quadrupole field complicate the optical pumping dynamics.

\begin{figure}[t]
\includegraphics[width=\linewidth]{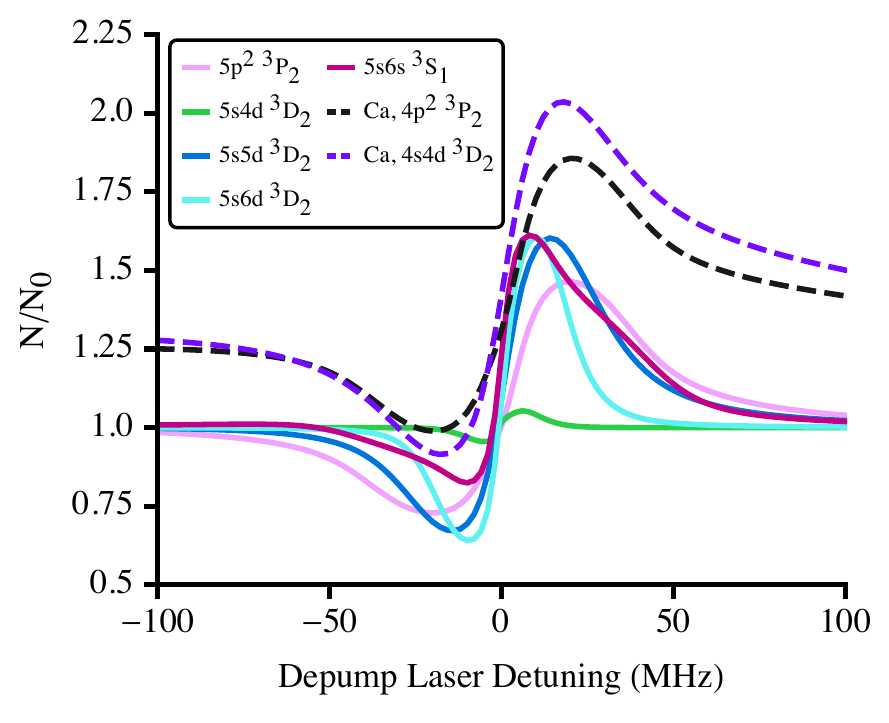}
\caption{(Color online) Simulations of several depumping schemes for bosonic strontium (solid lines) and calcium (dashed lines), which we label with the depumping transition excited state (the lower state is always $5s5p\>^{3}P_{1}$ for Sr and $4s4p\>^{3}P_{1}$ for Ca). In all simulations, we use our $^{88}\text{Sr}$ values for $r_{\text{bMOT}}$, $t_{\text{load}}$, $s_{461}$, $\Delta_{461}$, $\mathcal{F}_{^{1}S_{0}}$, and $\beta$. The depumper saturation parameter is $s_{depump}=1$ except for the simulation of $5s4d\>^{3}D_{2}$, where $s_{depump}=2000$, which requires much higher saturation due to the narrow transition linewidth. For the simulation of $5s6s\>^{3}S_{1}$, $\Delta_{679}$ is set to its optimal value. The Sr $5s5d\:^{3}D_{2}$, Sr $5s6d\:^{3}D_{2}$, Sr $5p^{2}\>^{3}P_{2}$, Ca $4p^{2}\>^{3}P_{2}$, and Ca $4s4d\>^{3}D_{2}$ states may indirectly decay to $^{3}P_{0}$ via intermediate states outside of the $^{3}P$ manifold. The model ignores these processes, but atoms decaying into $^{3}P_{0}$ can be recovered using $e.g.$ a 679-nm laser (for Sr). The apparent offset of the enhancement for Ca is a gaussian pedestal with a full width at half maximum of approximately $500$~MHz. The linewidths necessary for these simulations can be found in~\cite{Mickelson2009, Sansonetti2010,Werij1992,NIST_ASD}.}
\label{SchmSim}
\end{figure}

The choice of a $J\,{=}\,1\rightarrow J'\,{=}\,1$ transition as our depumping line potentially limits the trap loading enhancement, since the 688-nm line has position-dependent dark states and small CG overlap with $|\term{3}{P}{2},m_{J}=2\rangle$. Furthermore, this transition requires a secondary laser to depopulate the $\term{3}{P}{0}$ state. Many repumping strategies exist for strontium and each of these possesses a nearby depumping resonance~\cite{Stellmer2013,Mickelson2009,Stellmer2014,Stellmer2014a}. We assess the relative merit of the various schemes by simulating them with optimum parameters (Figure~\ref{SchmSim}). The $5s5p\>\term{3}{P}{1}\rightarrow 5s5d\,\term{3}{D}{2}$ line at 487~nm and the $5s5p\>\term{3}{P}{1}\rightarrow 5s6d\,\term{3}{D}{2}$ line at 397~nm have similar performance to the 688-nm line. All other transitions for which linewidth data are available give less enhancement. For the $5s5p\,\term{3}{P}{1}\rightarrow5p^{2}\,\term{3}{P}{2}$ transition, unfavorable relative Land\'{e} $g$-factors between the excited state and $\term{3}{P}{1}$ marginally reduce the trap loading improvement. The linewidth of the $5s5p\,\term{3}{P}{1}\rightarrow5s4d\,\term{3}{D}{2}$ transition is too narrow for efficient optical pumping at bMOT temperatures.

We investigate the utility of the depumping scheme for other AE atoms. For Cd, Hg, Yb, Be, and Mg, the $\term{1}{D}{2}$ state lies above the $\term{1}{P}{1}$ state, so efficient continuous loading of the metastable reservoir does not occur~\cite{Stellmer2013,NIST_ASD}. Direct pumping to the magnetically trapped state is possible for these atoms~\cite{Pandey2010}. The $\term{1}{D}{2}{:}\term{1}{S}{0}$ branching ratio in Ba and Ra is very large, which means that cooling to temperatures below the magnetic trap depth may not be possible without repumping~\cite{Guest2007,De2009}. The level structure of calcium combines several features that make depumping more effective than in strontium (see Figure~\ref{SchmSim}). The $\term{3}{P}{2}{:}\term{3}{P}{1}$ branching ratio is ${\approx}\,1{:}3$ and, more importantly, the $\term{1}{D}{2}\rightarrow\term{1}{S}{0}$ quadrupole transition linewidth is comparable to $\term{1}{D}{2}\rightarrow\term{3}{P}{J}$ decay rates. Ca can be trapped in a MOT operating on $\term{3}{P}{2}\rightarrow\term{3}{D}{3}$ transition~\cite{Grunert2002}, the loading of which could also benefit from this depumping technique. The loading enhancement for MOTs does not benefit from the detuning-dependent asymmetry seen for magnetic trap loading, which limits the simulated improvement in Ca to ${\approx}50~\%$.

\section{\label{con}Conclusions}

We have demonstrated that the 688-nm transition can be used to reduce cycle time and increase atom number in ultracold strontium experiments. For the bosonic isotopes, applying both a 688-nm and a 679-nm laser to the bMOT increases atom number in the metastable reservoir by up to $65~\%$ regardless of loading time. If an experiment requires large atom number relative to experimental limits, the trap loading time can be reduced by a factor of three or better. The enhancement is less for $\sr{87}$ due to complications arising from hyperfine structure and smaller Land\'{e} $g$-factors for the $\term{3}{P}{2}$ state. If a second frequency component to simultaneously pump $\ket{F\,{=}\,11/2}\rightarrow \ket{F'\,{=}\,11/2}$ and $\ket{F\,{=}\,9/2}\rightarrow \ket{F'\,{=}\,11/2}$ were added to the depumping beam, we believe performance comparable to the bosonic isotopes would be achievable. This improvement might make the depumping technique a useful method to reduce dead time in $\sr{87}$ atomic clocks~\cite{Nicholson2015}.

Comparison with a one-dimensional rate equation model shows that our results for the bosons are consist with expectations. The initial prediction of $3\times$ increased atom number, based on the branching ratio from $\term{1}{D}{2}$ into $\term{3}{P}{2}$ and $\term{3}{P}{1}$, is not feasible. Simulations of alternative enhancement schemes indicate that pumping on either the $5s5p\>\term{3}{P}{1}\rightarrow 5s5d\,\term{3}{D}{2}$ transition or the $5s5p\>\term{3}{P}{1}\rightarrow 5s6d\,\term{3}{D}{2}$ transition, which are also accessible with diode lasers, offers similar performance to the approach pursued in this work. Regardless of the exact implementation, the trap loading enhancement scheme can substantially increase atom number independent of the bMOT loading rate or vacuum lifetime. We expect that this method will be helpful for experiments benefitting from high atom number or faster cycle times.

The authors thank J.A. Pechkis for his work on the experimental apparatus. D.S. Barker acknowledges support from the NIST-ARRA Fellowship Program. This work was partially supported by ONR and the NSF through the PFC at the JQI.

\bibliography{Depump-Paper,Depump-Paper2}

\end{document}